\documentclass[twocolumn,showpacs,amsmath,amssymb]{revtex4}
\usepackage{graphicx}
\usepackage{dcolumn}
\usepackage{bm}

\def\stacksymbols #1#2#3#4{\def\theguybelow{#2}
    \def\verticalposition{\lower#3pt}
    \def\spacingwithinsymbol{\baselineskip0pt\lineskip#4pt}
    \mathrel{\mathpalette\intermediary#1}}
\def\intermediary#1#2{\verticalposition\vbox{\spacingwithinsymbol
    \everycr={}\tabskip0pt
    \halign{$\mathsurround0pt#1\hfil##\hfil$\crcr#2\crcr
        \theguybelow\crcr}}}

\begin{document}
\title{A right-handed isotropic medium with a negative refractive index}
\author{Yi-Fan Chen}
\email[Corresponding author: ]{yc245@cornell.edu}
\affiliation{Department of Applied and Engineering Physics,
Cornell University, Ithaca, NY 14853, USA}
\author{Peer Fischer}
\affiliation{Department of Applied and Engineering Physics,
Cornell University, Ithaca, NY 14853, USA}
\author{Frank W. Wise}
\affiliation{Department of Applied and Engineering Physics,
Cornell University, Ithaca, NY 14853, USA}

\date{\today}
\begin{abstract}
The sign of the refractive index of any medium is soley determined
by the requirement that the propagation of an electromagnetic wave
obeys Einstein causality. Our analysis shows that this requirement
predicts that the real part of the refractive index may be
negative in an isotropic medium even if the electric permittivity
and the magnetic permeability are both positive. Such a system may
be a route to negative index media at optical frequencies. We also
demonstrate that the refractive index may be positive in
left-handed media that contain two molecular species where one is
in its excited state.
\end{abstract}

\pacs{41.20.Jb,78.20.Ci,42.25.Fx} \maketitle
\begin{figure}[ht]
\centerline{\scalebox{1}{\includegraphics{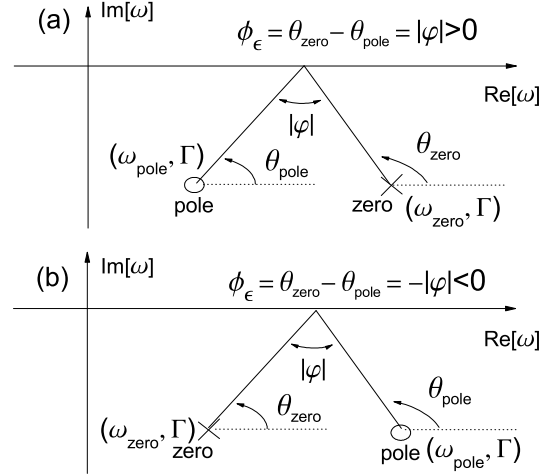}}}
\caption{\label{lorentz}The zero-pole pairs generated by (a) a
non-inverted Lorentz oscillator ($F>0$) and (b) an inverted
Lorentz oscillator ($F<0$). Only the poles and zeros for
$\mathrm{Re}[\omega]>0$ are indicated, as the the structures are
symmetric with respect to $\mathrm{Re}[\omega]=0$.}
\end{figure}
\begin{figure}[ht]
\centerline{\scalebox{1}{\includegraphics{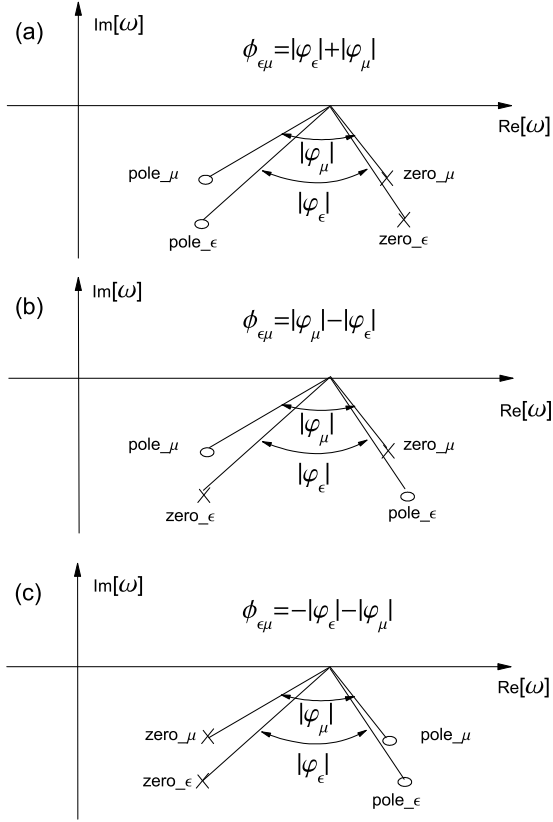}}}
\caption{\label{material} Three of the possible four different
types of zero-pole pairs are shown for LHM: (a) $F>0$ and $G>0$,
(b) $F<0$ and $G>0$ (c) $F<0$ and $G<0$. The one not show here is
a trivial case one can get by exchanging the role of $\epsilon$
and $\mu$ in (b).}
\end{figure}
\begin{figure}[ht]
\centerline{\scalebox{1}{\includegraphics{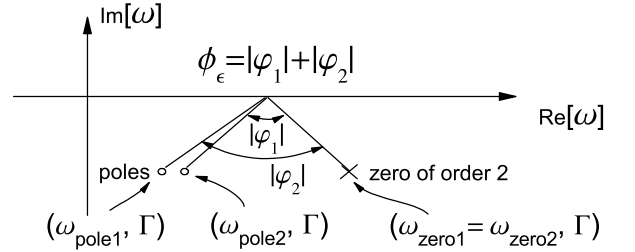}}}
\caption{\label{rhmindex} The zero-pole structures for a RHM
system with negative refractive index.}
\end{figure}
\begin{figure}[t]
\centerline{\scalebox{1}{\includegraphics{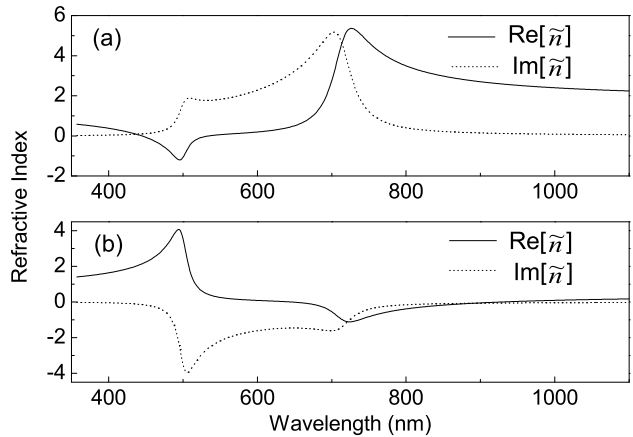}}}
\caption{\label{examples}Shown are the real and imaginary parts of
the refractive index. See text for details.}
\end{figure}

In 1968 Veselago
\cite{veselago} considered the electrodynamic properties of
isotropic media where the real part of the electric permittivity
$\epsilon$ and the real part of the magnetic permeability $\mu$
are simultaneously negative. Veselago showed that if
$\epsilon,\mu<0$ then the electric field $\mathbf{E}$, the
magnetizing field $\mathbf{H}$ and the wave-vector $\mathbf{k}$
form a left-handed orthogonal set, contrary to all known
naturally-occurring materials where the triplet of these vectors
is right-handed. Media with both negative electric permittivity
and magnetic permeability are referred to as left-handed materials
(LHM) at the frequencies for which
$\epsilon,\mu<0$. \\
\indent A consequence of simultaneously negative $\epsilon$ and
$\mu$ is that the Poynting vector
$\mathbf{S}=\mathbf{E}\times\mathbf{H}$ and the wave-vector
$\mathbf{k}=(\omega/|\mathbf{E}|^2) \mathbf{E}\times\mathbf{B}$
point in opposite directions for a monochromatic plane wave with
angular frequency $\omega$, as here the direction of the magnetic
field necessarily opposes that of the magnetizing field,
$\mathbf{B}=\mu\mathbf{H}$ \cite{veselago}. Veselago argued that
the direction of the energy flow ($\mathbf{S}$) must point away
from its source and thereby reached the surprising conclusion that
in LHM the wave-vector points toward the source \cite{veselago}.
This in turn lead to the prediction that LHM exhibit a negative
refractive index, as well as reversed Doppler and Cherenkov
effects \cite{veselago}.
\\ \indent No naturally-occurring
isotropic material is known to have $\epsilon,\mu<0$ at the same
frequency, since the underlying polarizabilities and
magnetizabilities, in general, exhibit different frequency
responses (resonances). Pendry {\it et al.} thus
\cite{pendrymicro} suggested that structures containing metal
strips and split-ring resonators could be engineered such that
both $\epsilon$ and $\mu$ are negative at microwave frequencies.
Shelby {\it et al.} \cite{shelby} subsequently reported the
observation of negative refraction at $10.5$ GHz in a left-handed
metamaterial. There are to date no experimental reports of
homogeneous, isotropic media that exhibit a negative refractive
index at optical frequencies. We note that neither photonic
crystals \cite{photonic1,photonic2}, nor birefringent crystal
assemblies \cite{zhang} are isotropic and can therefore not be
characterized by a single, scalar refractive index.
\\ \indent In
this Letter we consider several cases in which negative refraction
may be achieved in isotropic and homogeneous media. Some aspects
of the connection between Einstein causality and the sign of the
refractive index have been discussed previously
\cite{DSandNK,MandChiao}. Here we show that a standard analysis of
causal wave-propagation, as first discussed by Sommerfeld and
Brillouin \cite{brillouin}, reveals that the real part of the
refractive index, Re$[\widetilde{n}(\omega)]$, may in principle be
of either sign for a LHM. Remarkably, a negative refractive index,
Re$[\widetilde{n}(\omega)]<0$, may be observed even when both
$\epsilon,\mu>0$ (i.e.~a right-handed medium (RHM)). This may be
particularly promising for the observation of negative refraction
and associated phenomena \cite{veselago,pendry} at optical
frequencies in isotropic media.
\\ \indent We begin by writing the
wave equations for a homogeneous isotropic linear medium in a
source-free region as
\begin{equation}\label{laplace}
\left\{\begin{array}{r@{\quad=\quad}l} (\nabla)^2
\widetilde{\mathbf{E}}(\mathbf{r},\omega)&
-\omega^2\widetilde{\epsilon}(\omega)
\widetilde{\mu}(\omega)\widetilde{\mathbf{E}}(\mathbf{r},\omega) \\
(\nabla)^2 \widetilde{\mathbf{B}}(\mathbf{r},\omega)&
-\omega^2\widetilde{\epsilon}(\omega)
\widetilde{\mu}(\omega)\widetilde{\mathbf{B}}(\mathbf{r},\omega) .
\end{array}\right.
\end{equation}
$\widetilde{\mathbf{E}}(\mathbf{r},\omega)$ and
$\widetilde{\mathbf{B}}(\mathbf{r},\omega)$ are the complex
Fourier transforms of the corresponding real fields
$\mathbf{E}(\mathbf{r},t)$ and $\mathbf{B}(\mathbf{r},t)$, where
the Fourier transforms in (\ref{laplace}) are defined as
\begin{equation}\label{fourierdefine}
\widetilde{f}(\omega)=\frac{1}{\sqrt{2\pi}}\int_{-\infty}^{\infty}f(t)e^{i\omega
t} dt.
\end{equation}
The vector equations have a common form for each Cartesian
component and can be simplified to
\begin{equation}\label{laplace2}
\frac{\partial^2 \widetilde{g}(z,\omega)}{\partial z^2}=
-\omega^2\widetilde{\epsilon}(\omega)
\widetilde{\mu}(\omega)\widetilde{g}(z,\omega) ,
\end{equation}
if we consider a general plane wave propagating in the $+z$
direction. The equation has the time-domain Green function
solutions
\begin{eqnarray}
\label{generalsoltime} \displaystyle g
(z,t)&=&
\frac{1}{\sqrt{2\pi}}\int_{-\infty}^{\infty}
\widetilde{g}
(z,\omega)e^{-i\omega t} d\omega
\nonumber \\
&=& \frac{1}{\sqrt{2\pi}}\int_{-\infty}^{\infty}e^{i\big(
\widetilde{k}(\omega)z-\omega t \big)} d\omega,
\end{eqnarray}
where
$\widetilde{k}(\omega)\equiv\omega\sqrt{\widetilde{\epsilon}(\omega)
\widetilde{\mu}(\omega)}\equiv \widetilde{n}(\omega)\omega/c$, and
$\widetilde{n}(\omega)$ is the complex refractive index. We note
that $\widetilde{k}(\omega)$ is in general not single-valued. Only
when all branch-cuts have been defined and when a certain
branch-cut has been chosen does $\widetilde{k}(\omega)$ become
analytic (see for instance \cite{arfken}) such that the sign of
the refractive index can be determined. The causal propagation of
an electromagnetic wave requires that the Green functions $g
(z,t)=0$ for $t< (z/c)$ (see also \cite{field}). Further, the
following conditions hold i) due to a material's finite bandwidth
$\widetilde{\epsilon}(\omega)\rightarrow\epsilon_0$ and
$\widetilde{\mu}(\omega)\rightarrow\mu_0$ as
$|\omega|\rightarrow\infty$, and ii) $\widetilde{k}(\omega)$ is
analytic in the upper-half of the complex $\omega$-plane. The
latter implies that not only the poles but also the zeros of
$\widetilde{\epsilon}(\omega)$ and $\widetilde{\mu}(\omega)$ must
lie outside the upper-half of the complex $\omega$-plane, such
that the material response is causal. This can be shown using
Titchmarsh's theorem \cite{arfken,titchmarsh}. The branch-cuts of
$\widetilde{k}(\omega)$ can then be chosen to lie outside the
upper-half of the complex $\omega$-plane, and (the principal)
branch of $\widetilde{k}(\omega)\rightarrow (\omega/c)$ can be
chosen to lie on the real $\omega$ axis as $|\omega|
\rightarrow\infty$.
\\ \indent It is then possible to evaluate the
integral in Eq.~(\ref{generalsoltime}) by closing the contour in
the upper-half of the complex $\omega$-plane, and it follows that
$g (z,t)=0$ for $t< (z/c)$, as is required by causality
\cite{brillouin, chiao, diener, milonni}.
\\ \indent We now
express the permittivity and the permeability in terms of Lorentz
oscillator models as their generality facilitates the discussion
of the refractive index for a variety of systems, such as atoms in
the gas phase, conductors near a plasmon resonance, or any medium
whose optical properties are directly related to the underlying
molecular polarizabilities and magnetizabilities. The electric
permittivity $\widetilde{\epsilon}(\omega)$ and the magnetic
permeablity $\widetilde{\mu}(\omega)$ may then be written as
\begin{eqnarray}\label{lorentz1}
\widetilde{\epsilon}(\omega)&=&\epsilon_0 \left( 1 +
\frac{F}{\omega_{\mathrm{pole}\_\epsilon}^2-(\omega+i\Gamma)^2}\right)
\nonumber \\
&=&\epsilon_0\frac{(\omega+i\Gamma)^2-\omega_{\mathrm{zero}\_\epsilon}^2}{(\omega+i\Gamma)^2-\omega_{\mathrm{pole}\_\epsilon}^2}
\\ \widetilde{\mu}(\omega)&=&\mu_0 \left( 1 +
\frac{G}{\omega_{\mathrm{pole}\_\mu}^2-(\omega+i\Gamma)^2}\right)
\nonumber\\ \label{lorentz2}
&=&\mu_0\frac{(\omega+i\Gamma)^2-\omega_{\mathrm{zero}\_\mu}^2}{(\omega+i\Gamma)^2-\omega_{\mathrm{pole}\_\mu}^2},
\end{eqnarray}
where
$\omega_{\mathrm{zero}\_\epsilon}^2=\omega_{\mathrm{pole}\_\epsilon}^2+
F$, and similarly
$\omega_{\mathrm{zero}\_\mu}^2=\omega_{\mathrm{pole}\_\mu}^2+ G$.
$F, G$ and $\Gamma$ are all taken to be real. A system in the
ground state  corresponds to $F,G>0$ and an inverted system has
$F,G<0$. The permittivity of the vacuum is $\epsilon_0$ and its
permability is $\mu_0$. $\Gamma$ is the half width at half maximum
of the Lorentzian spectrum ($\Gamma>0$).  It is seen that
eqns.~(\ref{lorentz1}) and (\ref{lorentz2}) satisfy the conditions
above.
\\ \indent Writing
$\widetilde{\epsilon}(\omega)=|\widetilde{\epsilon}(\omega)|e^{i\phi_{\epsilon}(\omega)}$,
it can be seen in Fig.~\ref{lorentz} that the structures of the
zeros and poles determine $\phi_{\epsilon}$. In the case of a
non-inverted Lorentz oscillator, the contribution from the
zero-pole pair yields a positive $\phi_{\epsilon}$, whereas for
the inverted system, $\phi_{\epsilon}$ is negative.
For $\widetilde{\epsilon}\widetilde{\mu} = |\widetilde{\epsilon}|
|\widetilde{\mu}|e^{i(\phi_{\epsilon \mu})}$, the angle
$\phi_{\epsilon\mu}=\phi_{\epsilon}+\phi_{\mu}$ is thus determined
by the zeros and poles of both $\widetilde{\epsilon}$ and
$\widetilde{\mu}$, as is shown in Fig.~\ref{material}. It then
follows that the refractive index
$\widetilde{n}=|\widetilde{n}|e^{i\phi_{n}}=c\sqrt{\widetilde{\epsilon}\widetilde{\mu}}=
c\sqrt{|\widetilde{\epsilon}||\widetilde{\mu}|}e^{i(\phi_{\epsilon}+\phi_{\mu})/2}$
is uniquely determined. For propagating waves only the sign of its
real part is of interest.
\\ \noindent {\bf LHM with a positive refractive index:} For the
subsequent discussion we consider the cases for which the sign of
$F$ and $G$ in equations (\ref{lorentz1}) and (\ref{lorentz2}) can
independently vary. We suppose that this may, at least in
principle, be achieved in a multi-component system or in certain
multi-level systems. Considering a two-component system, we assume
that in a certain frequency range the electric permittivity is
dominated by one molecular species whereas the magnetic
permeability is dominated by the other species. If only one of the
species is inverted, then we may have $F$ and $G$ of opposite
sign. Similarly, in an inverted one-component multi-level system
there may be nearby lying states that have electric-dipole
allowed, but magnetic-dipole forbidden transitions, such that the
polarizability and magnetizability, which respectively underlie
$\widetilde{\epsilon}(\omega)$ and $\widetilde{\mu}(\omega)$, are
of opposite sign. If for these systems we describe
$\widetilde{\epsilon}(\omega)$ and $\widetilde{\mu}(\omega)$ by an
effective Lorentz oscillator model, then there are the following
possible combinations: $F,G>0$ (Fig.~\ref{material}(a)), $F$ and
$G$ of opposite sign (Fig.~\ref{material}(b)), only one case is
shown), and $F,G<0$ (Fig.~\ref{material}(c)).
\\ \indent A
left-handed medium requires $F$ in Eqn.~(\ref{lorentz1}) and $G$
in Eqn.~(\ref{lorentz2}) to be large compared to $\Gamma$, such
that $|\varphi_{\epsilon}|$ and $|\varphi_{\mu}|$ exceed
$(\pi/2)$, and such that both
$\mathrm{Re}[\widetilde{\epsilon}(\omega)]$ and
$\mathrm{Re}[\widetilde{\mu}(\omega)]$ are negative for some range
of $\omega$. However, since
$\phi_n=(\phi_{\epsilon\mu}/2)=(\phi_{\epsilon}+\phi_{\mu})/2$ and
since $\phi_{\epsilon}$ and $\phi_{\mu}$ can be of either sign,
$|\phi_n|$ can be smaller than $(\pi/2)$ depending on the specific
form of the zero-pole pair structure. Hence, a LHM does not
necessarily need to have a negative refractive index. Table
\ref{tab1} lists the four possible combinations. To our knowledge
only the passive LHM (Type I) has been discussed previously.
\begin{table}[ht]
\begin{tabular}{|c|c|c|c|}\hline
\rule[0mm]{0mm}{3.5mm} LHM & $\phi_{\epsilon}$ & $\phi_{\mu}$ &
$\mathrm{Re}[\widetilde{n}]$
\\ \hline\hline\rule[0mm]{0mm}{3.5mm}
Type I:& & &\  \\
$F>0$ & $(\pi/2)<\phi_{\epsilon}<\pi$ & $(\pi/2)<\phi_{\mu}<\pi$ & $<0$\\
$G>0$ & & &\
\\ \hline\rule[0mm]{0mm}{3.5mm}
Type II:& & &\  \\
$F<0$ & $-\pi<\phi_{\epsilon}<-(\pi/2)$ & $(\pi/2)<\phi_{\mu}<\pi$ & $>0$\\
$G>0$ & & &\
\\ \hline\rule[0mm]{0mm}{3.5mm}
Type III:& & &\  \\
$F>0$ & $(\pi/2)<\phi_{\epsilon}<\pi$ & $-\pi<\phi_{\mu}<-(\pi/2)$ & $>0$\\
$G<0$ & & &\
\\ \hline\rule[0mm]{0mm}{3.5mm}
Type IV:& & &\  \\
$F<0$ & $-\pi<\phi_{\epsilon}<-(\pi/2)$ & $-\pi<\phi_{\mu}<-(\pi/2)$ & $<0$\\
$G<0$ & & &\
\\ \hline
\end{tabular}
\caption{\label{tab1}Different LHM where $F$ and $G$ are defined
in eqns. (\ref{lorentz1}) and (\ref{lorentz2}), and where
$\phi_n=(\phi_{\epsilon}+\phi_{\mu})/2$.}
\end{table}
\\ \noindent {\bf RHM with a negative refractive index:} Now, we show
that even when both $\epsilon$ and $\mu$ are positive, i.e. a
right-handed medium (RHM), the structures of the zeros and poles
may in certain cases give rise to a negative refractive index.
Remarkably, this can happen for a non-magnetic system and is
therefore a potential route to negative index media at optical
frequencies. We consider a two-component system with $\mu=\mu_0$
and with
\begin{eqnarray}\label{rhmnegative}
\widetilde{\epsilon}(\omega)&=&\epsilon_0(1+\frac{\alpha}{\omega_{\mathrm{pole1}}^2-(\omega+i\Gamma)^2}
+\frac{\beta}{\omega_{\mathrm{pole2}}^2-(\omega+i\Gamma)^2})
\nonumber \\
&=&\epsilon_0\frac{((\omega+i\Gamma)^2-\omega_{\mathrm{zero1}}^2)((\omega+i\Gamma)^2-\omega_{\mathrm{zero2}}^2)}
{((\omega+i\Gamma)^2-\omega_{\mathrm{pole1}}^2)((\omega+i\Gamma)^2-\omega_{\mathrm{pole2}}^2)}.
\end{eqnarray}
We further assume that
$\omega_{\mathrm{pole2}}>\omega_{\mathrm{pole1}}$.
$\omega_{\mathrm{zero1}}$ and $\omega_{\mathrm{zero2}}$ depend on
$\alpha$ and $\beta$. Generally, $\alpha$ and $\beta$ are
independent of each other, however, to simplify the discussion we
consider the case for which
\begin{eqnarray}\label{alphabeta1}
\alpha &=&
\alpha_0(\omega_{\mathrm{pole2}}^2-\omega_{\mathrm{pole1}}^2)>0,
\nonumber \\
\beta &=& -(\sqrt{\alpha_0}\mp
1)^2(\omega_{\mathrm{pole2}}^2-\omega_{\mathrm{pole1}}^2)<0,
\end{eqnarray}
so that the two zeros are equal,
\begin{eqnarray}\label{alphabeta2}
\omega_{\mathrm{zero1}}^2 =
\omega_{\mathrm{zero2}}^2=\omega_{\mathrm{pole1}}^2\pm\sqrt{\alpha_0}(\omega_{\mathrm{pole2}}^2-\omega_{\mathrm{pole1}}^2),
\end{eqnarray}
where $\alpha_0$ is a positive real number. The poles and zeros
for the upper sign in Eqn.~(\ref{alphabeta1}) and
Eqn.~(\ref{alphabeta2}) are shown in Fig.~\ref{rhmindex}.
Provided that $\alpha_0$ is large enough, one may have
$(3\pi/4)<|\varphi_1|<\pi$ and $(3\pi/4)<|\varphi_2|<\pi$, such
that $(3\pi/2)<\phi_\epsilon<2\pi$. Hence, the real part of
$\widetilde{\epsilon}$ is positive. Given that $\mu=\mu_0$ is real
and positive, the system is a RHM. However, the refractive index
will have a negative real part, since
$(3\pi/4)<\phi_n=(\phi_\epsilon/2)<\pi$. Similarly, for the lower
sign in Eqn.~(\ref{alphabeta1}) and Eqn.~(\ref{alphabeta2}) (not
shown in the figure), we can have $-2\pi<\phi_\epsilon<-(3\pi/2)$.
Hence, the real part of $\widetilde{\epsilon}$ is again positive
and the system is also a RHM. The refractive index will also have
a negative real part, since
$-\pi<\phi_n=(\phi_\epsilon/2)<-(3\pi/4)$. It is interesting to
note that for this system the imaginary part of the refractive
index is now negative. This indicates that while the system with
the upper sign in Eqn.~(\ref{alphabeta1}) and
Eqn.~(\ref{alphabeta2}) is lossy, the system with the lower sign
has gain.
\\ \indent Numerical estimates show that for certain frequency ranges
RHM with negative refractive index may be realized experimentally.
We considered gas-phase as well as condensed-phase media, and here
we provide two specific examples using values typical of condensed
molecular media. In these examples the number densities are
somewhat high, however, the results show that it is plausible that
such systems may be realized.
\\ \indent $\alpha$ and $\beta$ are
given by
\begin{equation}\label{realnumber}
\alpha = \frac{2N_1 \overline{p}_1^2
\omega_{\mathrm{pole1}}}{\epsilon_0\hbar} \qquad \mathrm{and}
\qquad \beta = \frac{2N_2 \overline{p}_2^2
\omega_{\mathrm{pole2}}}{\epsilon_0\hbar},
\end{equation}
where $\overline{p}_1$ and $\overline{p}_2$ are the transition
dipole moments, and $N_1$ and $N_2$ the number densities of
molecular species 1 and 2, respectively. We now assume that
species 1 is in its ground state and has $N_1=3\times 10^{28}$
m$^{-3}$ with a resonance at $\omega_{\mathrm{pole1}}=14,000$
cm$^{-1}$. Further, we take $\overline{p}_1=\overline{p}_2=3$
Debye, and $\Gamma=400$ cm$^{-1}$. If species 2 is in the excited
state at $\omega_{\mathrm{pole2}}=20,000$ cm$^{-1}$ with
$N_2=2.52\times 10^{27}$ m$^{-3}$, then the real part of the
refractive index becomes negative between $445$ and $530$ nm. For
the wavelengths between $455$ and $495$ nm the material is
moreover right-handed, i.e., the real part of the refractive index
dominates, and reaches $-1.2$ in this wavelength range. The
parameters used above are based on the upper sign in
Eqn.~(\ref{alphabeta1}) and Eqn.~(\ref{alphabeta2}), and the
system is lossy for the frequency range where the real part of the
refractive index becomes negative, as can be seen in
Fig.~\ref{examples}(a).
\\ \indent Similarly, we can envision a
system with gain by considering the lower sign case in
Eqn.~(\ref{alphabeta1}) and Eqn.~(\ref{alphabeta2}). Assuming,
$N_1=2\times 10^{27}$ m$^{-3}$ for the non-inverted species,
$N_2=1.75\times 10^{28}$ for the inverted species,
$\omega_{\mathrm{pole1}}=14,000$ cm$^{-1}$,
$\omega_{\mathrm{pole2}}=20,000$ cm$^{-1}$, $\Gamma=400$
cm$^{-1}$, and taking $\overline{p}_1=\overline{p}_2=3$ Debye, we
find that the real part of the refractive index becomes negative
(for the wavelength range $650$ to $920$ nm), and within the range
of $720$ to $915$ nm, the material is right-handed and its
refractive index reaches $-1.1$ (see Fig.~\ref{examples}(b)).
However, now the imaginary part of the index is also negative,
i.e. the system has gain which may facilitate observation of the
effect. It is also clear that a negative refractive index can
appear when the material is not right-handed, i.e. when the real
part of the permittivity is negative.
\\ \indent We have considered the connection between Einstein causality and
the refractive index based on Sommerfeld and Brillouin theory of
causal wave propagation \cite{brillouin,chiao,diener,milonni}. The
refractive index is uniquely determined for any given $\epsilon$
and $\mu$. Considering both inverted systems and two-component
systems, we have shown that a LHM does not need to have a negative
refractive index. We predict the existence of non-magnetic RHM
systems where the real part of the refractive index is negative.
The latter suggests a new class of negative refractive index media
and may be particularly promising for experiments at optical
frequencies.
\begin{acknowledgments}
This work was supported by the National Science Foundation
(PHY-0099564, CHE-0095056). P.F. is grateful for a grant from the
Eppley Foundation for Research.
\end{acknowledgments}


\begin{references}
\bibitem{field} The field $\psi(z,t)$ due to an arbitrary source $s(t)$ located
at $z=0$ is given by
\begin{equation}\label{finalsol2}
\psi(z,t)= \big(\frac{1}{\sqrt{2\pi}}\big)\int_{-\infty}^{\infty}
s(t') g^{
}(z,t-t') dt'.\nonumber
\end{equation}
Given that $g (z,t)=0$ for $t< (z/c)$, the field $\psi(z,t)$
depends only on the source $s(t)$ for time later than $t>(z/c)$,
such that the "onset of the disturbance" (c.f. \cite{brillouin})
propagates at exactly the speed $c$.
\end{references}
\end{document}